\documentclass[twocolumn,aps,showpacs,floatfix,prc]{revtex4-2}
\usepackage[dvips]{epsfig}
\usepackage{amsmath,bm}
\usepackage{xcolor}[dvips]

\begin{document}
%\draft

\title{Revisiting the Lithium abundance problem in Big-Bang nucleosynthesis}

\author{ Vinay Singh$^{1}$, Debasis Bhowmick$^{2}$ and D. N. Basu$^{3}$ }

\affiliation{Variable Energy Cyclotron Centre, 1/AF Bidhan Nagar, Kolkata 700 064, India }

\email[E-mail 1: ]{vsingh@vecc.gov.in}
\email[E-mail 2: ]{dbhowmick@vecc.gov.in}
\email[E-mail 3: ]{dnb@vecc.gov.in}

\date{\today }

\begin{abstract}

    One of the three testaments in favor of the big bang theory is the prediction of the primordial elemental abundances in the big-bang nucleosynthesis (BBN). The Standard BBN is a parameter-free theory due to the precise knowledge of the baryon-to-photon ratio of the Universe obtained from studies of the anisotropies of cosmic microwave background radiation. Although the computed abundances of light elements during primordial nucleosynthesis and those determined from observations are in good agreement throughout a range of nine orders of magnitude, there is still a disparity of $^7$Li abundance overestimated  by a factor of $\sim 2.5$ when calculated theoretically. The number of light neutrino flavors, the neutron lifetime and the baryon-to-photon ratio in addition to the astrophysical nuclear reaction rates determine the primordial abundances. We previously looked into the impact of updating baryon-to-photon ratio and neutron lifetime and changing quite a few reaction rates on the yields of light element abundances in BBN. In this work, calculations are performed using new reaction rates for $^3$H(p,$\gamma$)$^4$He, $^6$Li(p,$\gamma$)$^7$Be, $^7$Be(p,$\gamma$)$^8$B, $^{13}$N(p,$\gamma$)$^{14}$O, $^7$Li(n,$\gamma$)$^8$Li and $^{11}$B(n,$\gamma$)$^{12}$B along with the latest measured value of neutron lifetime. We observe from theoretical calculations that these changes result in improvement by causing further reduction in the abundance of $^7$Li than calculated earlier. 
    
\vspace{0.2cm}
\end{abstract}

\maketitle

\noindent
\section{Introduction}
\label{section1}

    The three hallmarks of the big bang concept are the Hubble expansion of the Universe, the Cosmic Microwave Background Radiation (CMBR), and the big bang nucleosynthesis (BBN). A substantial amount of observational research backs them up. Only a few seconds after the great bang \cite{Ho64}, according to the BBN, which predicts the primordial abundances of the light elements like D, $^{3,4}$He and $^{6,7}$Li, did the universe begin to expand at a very rapid rate, permitting only the synthesis of the lightest nuclides. Together with these stable nuclei, the BBN also resulted in the production of unstable or radioactive isotopes including tritium or ${^3}$H and $^{7,8}$Be. One of the stable isotopes was created from these unstable isotopes by decay or nuclear fusion. The density and temperature of the universe then dropped below those needed for nuclear fusion, preventing the formation of elements heavier than beryllium while allowing unburned light elements like deuterium to exist. This phenomenon lasted for only about seventeen minutes (during the period from three to about twenty minutes after the beginning of space expansion).
    
    Although there is considerable agreement between the estimated primordial abundances of D and $^{3,4}$He derived from observations and estimates of the primordial nucleosynthesis, those of $^{6,7}$Li are off by significant amounts. The predictions of the conventional BBN theory depend on three extra factors, namely the number of light neutrino flavors, the neutron lifetime and the baryon-to-photon ratio of Universe, in addition to the astrophysical nuclear reaction rates. The baryon-to-photon ratio of the universe was precisely extracted from the data collected by the Planck \cite{Planck,Planck1} and Wilkinson Microwave Anisotropy Probe [WMAP] \cite{WMAP,WMAP1} space missions. The conventional theory of the weak interaction provides the weak reaction rates necessary for n-p equilibrium. With neutron lifetime as the only direct experimental input, elemental abundances are computed \cite{Di82}. The recent experimental value of neutron lifetime 880.3$\pm$1.1 s \cite{Ol14} used in our previous work \cite{Si19} is further updated \cite{Wi11,Yo14} to $877.75\pm0.28_{stat}+0.22/-0.16_{syst}$ s \cite{Go21}, which affects  the generation of $^{4}$He \cite{Ma05}. The BBN model's sensitivity to various parameters and physics inputs has previously been studied \cite{No00,Cy02,Cy04,Cy08,Se04,Io09,Fu10}.

    The nuclear reaction rates $<\sigma v>$ in the reaction network computations, where $\sigma$ is the nuclear fusion cross section and $v$ is the relative velocity between the colliding nuclei, are the most crucial inputs for modeling the BBN and stellar evolution. Only experiments performed in laboratories can yield these low energy fusion cross sections and some of which are not yet measured \cite{No00,Cy02,Cy04,Cy08,Se04}. A Maxwellian velocity distribution, however, accurately describes $v$ for a given temperature $T$. The measured cross section values are affected by a number of variables, particularly the numerous approximations employed in theory to determine thermonuclear reaction rates. The description of synthesis of elements in the BBN or in stellar evolution depends on the Maxwellian-averaged thermonuclear reaction rates \cite{Fo88,An99}, which must be taken into account in the network computations.    
      
    Previously we studied the effects of modifying thirty-five reaction rates on light element abundance yields in BBN \cite{Mi12}. Subsequently, we have updated primordial abundances \cite{Si19} by using the improved values of neutron lifetime, baryon-to-photon ratio and modified reaction rates for d($^4$He,$\gamma$)$^6$Li, t($^4$He,$\gamma$)$^7$Li and $^3$He($^4$He,$\gamma$)$^7$Be \cite{Du17} in the temperature ranges up to 5$T_9$ (in units of $10^9$ K), which is used directly for estimating the formation of $^7$Li as a result of decay by electron capture. In this work, calculations are performed for estimating light element abundance yields as functions of temperature and evolution time by incorporating six new reaction rates for $^3$H(p,$\gamma$)$^4$He, $^6$Li(p,$\gamma$)$^7$Be, $^7$Be(p,$\gamma$)$^8$B, $^{13}$N(p,$\gamma$)$^{14}$O, $^7$Li(n,$\gamma$)$^8$Li and $^{11}$B(n,$\gamma$)$^{12}$B along with the most recent experimental neutron lifetime measurement value of  $877.75\pm0.28_{stat}+0.22/-0.16_{syst}$ s \cite{Go21}.

\noindent
\section{Big-bang nucleosynthesis reaction network}
\label{section2}

    The reactions which occurred during early universe can be classified into two groups, {\it viz} reactions that transform neutrons into protons and vice versa, such as n $+$ e$^+ \leftrightarrow$ p $+~\bar\nu_e$, p $+$ e$^- \leftrightarrow$ n $+~\nu_e$ as well as n $\leftrightarrow$ p $+$ e$^-+\bar\nu_e$ and the remaining reactions. While the second group is determined by numerous distinct nuclear cross section measurements, the first group can be defined in terms of the mean lifetime of neutron. Through the reaction p $+$ n $\leftrightarrow$ D $+~\gamma$, deuterium creation starts. The energy difference in this exothermic reaction is 2.2246 MeV, but because photons are 10$^9$ times more abundant than protons, the reaction does not start until the temperature of the expanding universe drops down to about 0.3 MeV, at which point the rate of photo-destruction is slower than the rate of deuterons being produced. Following the deuteron formation, other reactions create ${^4}$He nuclei, including D$~+ n \rightarrow {^3}$H$~+ \gamma$, ${^3}$H$~+~$p $\rightarrow {^4}$He $+$ $\gamma$, D $+$ p $\rightarrow {^3}$He $+~ \gamma$, ${^3}$He $+$ n $\rightarrow {^4}$He $+~\gamma$. Along with the ${^3}$H, both light helium ${^3}$He and regular helium ${^4}$He are created. Since the temperature has already dropped to 0.1 MeV and the helium nucleus is more tightly bonded than deuterons with a binding energy of 28.3 MeV, these interactions can only be one-way photo-reactions. The four reactions that also result in ${^3}$He and ${^4}$He are D $+$ D $\rightarrow {^3}$He $+$ n, D $+$ D $\rightarrow {^3}$H + p, ${^3}$He $+$ D $\rightarrow {^4}$He $+$ p, ${^3}$H $+$ D $\rightarrow {^4}$He $+$ n. These reactions often proceed more quickly as they do not associate the comparatively slow process of photon emission. Ultimately the temperature becomes so low that the electrostatic repulsion of the deuterons and other charged particles results the reactions to stop. When the processes come to an end, the deuteron to proton ratio is very modest and virtually inversely proportional to the combined density of protons and neutrons (to be specific, it goes approximately as the -1.6 power of density). Nearly all of the universe's neutrons end up in normal helium nuclei. About twenty five percent of the mass is converted into helium for a neutron:proton ratio of 1:7 at the time of deuteron production. Around 100 seconds after the big bang, deuterium reaches its peak and is then quickly swept up into helium nuclei. A minor abundance of ${^7}$Li is produced from the big bang when a very small number of helium nuclei combine into heavier nuclei. No ${^3}$H has survived to the present day due to the decay of ${^3}$H into ${^3}$He, which has a half-life of twelve years, and the decay of ${^7}$Be into ${^7}$Li, which has a half-life of around fifty-three days, also does not survive.         

    The nuclear reaction inputs to BBN take the form of thermal rates rather than cross sections $\sigma$. Nuclear reaction cross sections are averaged over a Maxwell-Boltzmann distribution of energies to determine the thermonuclear reaction rates. The following integral gives the Maxwellian-averaged thermonuclear reaction rate per particle pair $<\sigma v>$ at temperature $T$ \cite{Bo08}:

\begin{equation}
 <\sigma v> = \Big[\frac{8}{\pi\mu (k_B T)^3 } \Big]^{1/2} \int \sigma(E) E \exp(-E/k_B T) dE,
\label{seqn4}
\end{equation}
\noindent
where $E$ is the energy in the centre-of-mass system, $v$ is the relative velocity and $\mu$ is the reduced mass of the interacting nuclei. At energies much below Coulomb barrier, the classical turning point is much larger than the nuclear radius and barrier penetrability approximately behaves as $\exp(-2\pi\zeta)$ so that the charge induced cross section can be factorized into

\begin{equation}
 \sigma(E) = \frac{S(E)\exp(-2\pi\zeta)}{E}
\label{seqn5}
\end{equation}
\noindent
where $\zeta = \frac{Z_1Z_2e^2}{\hbar v}$ is the Sommerfeld parameter with $Z_1$ and $Z_2$ being the charges of the interacting nuclei in units of elementary charge $e$ and $S(E)$ is the astrophysical $S$-factor. The fact that the $S(E)$ factor is a smooth function of energy (excepting narrow resonances) facilitates extrapolation of measured cross sections down to very low energies of astrophysical domain. The resonant cross section $\sigma_r(E)$ in case of a narrow resonance can be approximated by a Breit-Wigner expression while the neutron induced reaction cross sections can be given by $\sigma(E)=\frac{R(E)}{v}$ \cite{Bl55} at low energies enabling extrapolation of the measured cross sections down to astrophysical energies, where $R(E)$, like the $S$-factor, is a slowly varying function of energy \cite{Mu10}.

\noindent
\section{The primordial abundances: Present observational status}
\label{section3}

    Stars also produce ${^4}$He after BBN. The primordial abundance of it is inferred from studies in the ionized hydrogen regions of compact blue galaxies. Galaxies are thought to be produced by the accumulation of these more primitive dwarf galaxies, according to a hierarchical structure creation paradigm. To account for stellar production, ${^4}$He abundance deduced from observations is extrapolated to zero, followed by atomic physics corrections. Aver et al. estimated its mass fraction to be $0.2449 \pm 0.0040$ \cite{Aver15} which was revised to $0.2453 \pm 0.0034$ \cite{Aver15} in the year 2020 with lower uncertainty than earlier estimate.  

    After BBN, deuterium can be eliminated throughout the evolution of stars. Observing a few cosmic clouds in the line of sight of far-off quasars at high redshift, primordial abundance of deuterium can be inferred. The recent reanalysis of prior data and additional observations by Cooke et al. \cite{Cooke14} resulted in a D/H relative abundance of $(2.53 \pm 0.04) \times 10^{-5}$ which was revised to $(2.527 \pm 0.030) \times 10^{-5}$ \cite{Cooke14} in the year 2018 with lower uncertainty than earlier estimate.

    In contrast to ${^4}$He, ${^3}$He is both created and destroyed in stars, making it difficult to predict how its abundance will change over time. The relative abundance of ${^3}$He is predicted to be $(<1.1 \pm 0.2) \times 10^{-5}$ \cite{Ba02} based on the observations in our Galaxy only due to the challenges faced in helium investigation and the low ${^3}$He/${^4}$He ratio.
    
    Three to twenty minutes after the start of space expansion, the BBN proceeded. As a result, the universe's temperature and density decreased below the level needed for nuclear fusion, which prevented the formation of materials heavier than beryllium while permitting the existence of unburned light elements like deuterium. The massive stars are where the heavy element nucleosynthesis occurs most frequently. These stars undergo supernova explosions and release materials rich in heavy metals into the interstellar medium as galaxies evolve. As a result, abundances of heavier elements in stars grows with time. Accordingly, age of star can be inferred from the measured abundance of metals (elements heavier than helium). Consequently, the metallicity is lower in the older stars. In order to derive the primordial abundances, observations of items with very low metallicity were made. Following BBN, $^7$Li can both be created (via spallation, AGB stars, and novae) and destroyed. (in the interior of stars). Since stars with masses less than the Sun have a lifespan that is longer than the age of the universe, very old stars can still be seen in the halo of our galaxy. Lithium can be seen at the surfaces of such low metallicity old stars and it was discovered that its abundance is astonishingly consistent regardless of metallicity as long as it is less than $\approx$0.1 of solar metallicity. This consistent plateau \cite{Sp82} in Li abundance was thought to correspond to the synthesis of $^7$Li in BBN. The plateau's thinness suggests that surface Li depletion may have been rather ineffective and it ought to have reflected the primordial value. The investigation by Sbordone et al. \cite{Sb10} estimates $^7$Li/H $=(1.58^{+0.35}_{-0.28}) \times 10^{-10}$.   

\noindent
\section{ Theoretical calculations }
\label{section4}

    Previously \cite{Mi12}, we modified thirty-five Maxwellian-averaged thermonuclear reaction rates from Caughlan et al. \cite{Fo88} and Smith et al. \cite{Sm93} used in the Kawano/Wagoner BBN code \cite{Wa67,Wa69,Ka92} by the compilations of Angulo et al. \cite{An99} and Descouvemont et al. \cite{An04} which were meant to supersede the earlier compilations and studied its effect on the primordial elemental abundance estimates. These rates were latest at that point of time, but there was a limitation in these rates as only some specific experimental values were adopted. Although the rate uncertainties were better defined in the Nollett and Burles \cite{No00} compilation than in Smith et al. \cite{Sm93} or Angulo et al. \cite{An99}, but the astrophysical $S$-factors of \cite{No00} were fitted by splines which had little physical justification \cite{An04}. Subsequently, we used the latest estimates of the baryon-to-photon ratio, the neutron lifetime, and the reaction rate of $^3$He($^4$He,$\gamma$)$^7$Be, which is used directly for calculating the creation of $^7$Li as a result of decay by electron capture \cite{Du17}. In the temperature ranges up to 5$T_9$, we have also used the most recent  reaction rates \cite{Du17} for t($^4$He,$\gamma$)$^7$Li and d($^4$He,$\gamma$)$^6$Li. In the present work, we have employed new reaction rates for $^3$H(p,$\gamma$)$^4$He, $^6$Li(p,$\gamma$)$^7$Be, $^7$Be(p,$\gamma$)$^8$B, $^{13}$N(p,$\gamma$)$^{14}$O, $^7$Li(n,$\gamma$)$^8$Li and $^{11}$B(n,$\gamma$)$^{12}$B and used the most recent experimentally measured neutron lifetime value of $877.75\pm0.28_{stat}+0.22/-0.16_{syst}$ s \cite{Go21}. We also contrasted findings of the current calculations with those of our earlier results \cite{Si19,Mi12} and with other more recent computations \cite{Coc14,Coc15,Cy16}.    
                
\noindent
\subsection{ Fundamental constants for primordial nucleosynthesis }
\label{subsection4a}

    The number of light neutrino flavors, the mean neutron lifetime and the baryon-to-photon ratio of the Universe are three characteristics that affect primordial abundances in addition to the astrophysical thermonuclear reaction rates. The number of light neutrino flavors $N_\nu$ is assumed to be 3.0 in the standard form. The measurements made by the Planck \cite{Planck,Planck1} and the WMAP \cite{WMAP,WMAP1}  space missions allowed for the precise determination of the baryon-to-photon ratio of the Universe, which is $\eta=6.0914\pm0.0438 \times 10^{-10}$. The present calculations employed the most recent experimental value for the mean neutron lifetime $\tau_n$, which is $877.75\pm0.28_{stat}+0.22/-0.16_{syst}$ s \cite{Go21}.

\begin{table*}[htbp]
\vspace{0.0cm}
\centering
\caption{\label{tab:table1} Yields at CMB-WMAP baryonic density ($\eta_{10}=6.0914 \pm 0.0438$ ~\cite{WMAP,WMAP1}).}
\begin{tabular}{ccccccl}
\hline
\hline
 &2012 \cite{Mi12}&2014 \cite{Coc14}&2015 \cite{Coc15}&2016 \cite{Cy16} &2019 \cite{Si19}&\\ \hline
 
${^4}$He&0.2479&0.2482$\pm$0.0003&0.2484$\pm$0.0002&0.2470&0.2467$\pm$0.0003 \\
D/H$~(\times 10^{-5})$&2.563&2.64$^{+0.08}_{-0.07}$&2.45$\pm$0.05&2.579 &2.623$\pm$0.031 \\
${^3}$He/H$~(\times 10^{-5})$&1.058&1.05$\pm$0.03&1.07$\pm$0.03&0.9996 &1.067$\pm$0.005\\
${^7}$Li/H$~(\times 10^{-10})$&5.019&4.94$^{+0.40}_{-0.38}$&5.61$\pm$0.26&4.648&4.447$\pm$0.067 \\
\hline
\hline

 &2020 \cite{Il20}&2021 \cite{Pi21}&This work &Observations&Ref. Year \\ \hline
${^4}$He&0.24709(18)&0.24721$\pm$0.00014&  0.2461$\pm$0.0002&0.2453$\pm$0.0034 &\cite{Aver15}~Aver et al. 2020 \\
D/H$~(\times 10^{-5})$&2.445(37)& 2.439$\pm$0.037&  2.620$\pm$0.031&2.527$\pm$0.030 &\cite{Cooke14}~Cooke et al. 2018  \\
${^3}$He/H$~(\times 10^{-5})$&1.041(18)&1.039$\pm$0.014&1.066$\pm$0.005&$<$1.1$\pm$0.2 &\cite{Ba02}~Bania et al. 2002 \\
${^7}$Li/H$~(\times 10^{-10})$&5.52(22)&5.464$\pm$0.220&  4.421$\pm$0.066 &1.58$^{+0.35}_{-0.28}$ &\cite{Sb10}~Sbordone et al. 2010 \\
 
\hline
\hline 
\end{tabular} 
\vspace{-0.5cm}
\end{table*}
\noindent     

\noindent
\subsection{ New thermonuclear reaction rates }
\label{subsection4b}

    The four lightest elements, ${^4}$He, D, ${^3}$He and ${^7}$Li, for which primordial abundances are predicted depends most sensitively on the twelve most significant nuclear processes. These are n$-$decay, p(n,$\gamma$)d, d(p,$\gamma){^3}$He, d(d,n)${^3}$He, d(d,p)t, ${^3}$He(n,p)t, t(d,n)${^4}$He, ${^3}$He(d,p)${^4}$He, ${^3}$He($\alpha,\gamma){^7}$Be, t($\alpha,\gamma){^7}$Li, ${^7}$Be(n,p)${^7}$Li and ${^7}$Li(p,$\alpha){^4}$He. The uncertainty in the projected yield of ${^7}$Li is immediately reflected in the uncertainties for the reactions reactions ${^3}$He $+$ ${^4}$He $\rightarrow {^7}$Be $+$ $\gamma$, ${^3}$H $+$ ${^4}$He $\rightarrow {^7}$Li $+$ $\gamma$ and p $+$ ${^7}$Li $\rightarrow {^4}$He $+$ ${^4}$He which we explored earlier \cite{Si19}. In the present work six new reaction rates for $^3$H(p,$\gamma$)$^4$He, $^6$Li(p,$\gamma$)$^7$Be, $^7$Be(p,$\gamma$)$^8$B, $^{13}$N(p,$\gamma$)$^{14}$O, $^7$Li(n,$\gamma$)$^8$Li and $^{11}$B(n,$\gamma$)$^{12}$B along with the latest measured value of mean neutron lifetime have been incorporated to re-investigate the problem of primordial elemental abundances in general and ${^7}$Li abundance in particular. Since only n$-$decay lifetime in the above list of new inputs figures in the twelve most important nuclear processes, it is unlikely to achieve any great improvement in primordial elemental abundance estimates.   

\begin{table*}[htbp]
\vspace{0.0cm}
\centering
\caption{\label{tab:table1} Variation of elemental abundances with neutron lifetime and reaction rates at CMB-WMAP baryonic density ($\eta_{10}=6.0914 \pm 0.0438$ ~\cite{WMAP,WMAP1}).}
\begin{tabular}{ccccl}
\hline
\hline
 &$\tau_n$(s) \cite{Ol14}&$\tau_n$(s) \cite{Ol14}&$\tau_n$(s) \cite{Go21}&   $\tau_n$(s) \cite{Go21}\\ \hline
&880.3$\pm$1.1&880.3$\pm$1.1&877.75$^{+0.28}_{-0.28}$$_{stat}$$^{+0.22}_{-0.16}$$_{syst}$&877.75$^{+0.28}_{-0.28}$$_{stat}$$^{+0.22}_{-0.16}$$_{syst}$ \\ \hline  
&Old rates&New rates&New rates&New rates$\pm$error \\ \hline 
${^4}$He&0.2467$\pm$0.0003&0.2467$\pm$0.0003&0.2461$\pm$0.0002&0.2461$\pm$0.0002$\pm<$ 4 significant digits \\

D/H~($\times 10^{-5}$)&2.623$\pm$0.031&2.623$\pm$0.031&2.620$\pm$0.031&2.620$\pm$0.031$\pm~~~<$ 4 significant digits \\

${^3}$He/H~($\times 10^{-5}$)&1.067$\pm$0.005&1.067$\pm$0.005&1.066$\pm$0.005&1.066$\pm$0.005$\pm~~~<$ 4 significant digits\\

${^7}$Li/H~($\times 10^{-10}$)&4.447$\pm$0.067&4.427$\pm$0.067&4.421$\pm$0.066&4.421$\pm$0.066$\pm$0.001 \\

\hline
\hline
\end{tabular} 
\vspace{-0.5cm}
\end{table*}
\noindent
    
     The refined variant of calculations for the astrophysical $S$-factor for the reactions mentioned above, are in better agreement with the previously available as well as the most recent experimental data and their predictive reliability have also been demonstrated. The new parametrization of the reaction rates for $^3$H(p,$\gamma$)$^4$He, $^6$Li(p,$\gamma$)$^7$Be, $^7$Be(p,$\gamma$)$^8$B, $^{13}$N(p,$\gamma$)$^{14}$O, $^7$Li(n,$\gamma$)$^8$Li and $^{11}$B(n,$\gamma$)$^{12}$B \cite{Du17a,Du22,Du19,Du20,Du21,Du20a} are given, respectively, by\\
     
\noindent
\vspace{0.25cm}      
$^3$H(p,$\gamma$)$^4$He \cite{Du17a} used for temperature range $\le 5T_9$       
\begin{eqnarray}
 N_A<\sigma v> =\frac{2.2182\times10^4}{T_9^{2/3}}\exp\Big(\frac{-4.3306}{T_9^{1/3}}\Big) \nonumber\\
      \times\Big[1.0+5.7852~T_9^{1/3}-11.374~T_9^{2/3}+12.059~T_9\Big] \nonumber\\
                -9.4143~T_9^{4/3}+47.332~T_9^{5/3},  
\label{seqn3}
\end{eqnarray}
\vspace{0.0cm}

\noindent
\vspace{0.25cm} 
$^6$Li(p,$\gamma$)$^7$Be \cite{Du22} used for temperature range $\le 10T_9$   
\begin{eqnarray}
N_A<\sigma v> =\frac{0.00319}{T_9^{1/3}}\exp\Big(\frac{-4.16292}{T_9^{2/3}}\Big) \nonumber\\
                 \times\Big[1.0+3721.884~T_9^{1/3}-544516.9~T_9^{2/3} \nonumber\\
                 +16401.8~T_9-8044.932~T_9^{4/3}+85197.34~T_9^{5/3}\Big] \nonumber\\
                 +924167.3~T_9^{2/3}\exp\Big(\frac{-8.36494}{T_9^{1/3}}\Big),  
\label{seqn4}
\end{eqnarray}
\vspace{0.0cm}

\noindent
\vspace{0.25cm} 
$^7$Be(p,$\gamma$)$^8$B \cite{Du19} used for temperature range $\le 10T_9$
\begin{eqnarray}
N_A<\sigma v> =\frac{81.15034}{T_9^{2/3}}\exp\Big(\frac{-10.3934}{T_9^{1/3}}\Big) \nonumber\\
                 \times\Big[1.0+20259.6~T_9^{1/3}-29831.13~T_9^{2/3} \nonumber\\
                 +7725.04~T_9+3133.587~T_9^{4/3}-1134.655~T_9^{5/3}\Big] \nonumber\\
                 +\frac{73867.4}{T_9^{1/2}}\exp\Big(\frac{-9.01236}{T_9^{1/2}}\Big),    
\label{seqn5}
\end{eqnarray}
\vspace{0.0cm}

\noindent   
\vspace{0.25cm} 
$^{13}$N(p,$\gamma$)$^{14}$O \cite{Du20} used for temperature range $\le 10T_9$    
\begin{eqnarray}
N_A<\sigma v> =\frac{4.68425}{T_9}\exp\Big(\frac{-5.5271}{T_9}\Big) 
    \times\Big[1.0+ \nonumber\\
    +72207.8~T_9^{1/3}-2.86832~T_9^{2/3}-17716.6~T_9^{4/3}  \nonumber\\
    -1155.274~T_9^{5/3}-1020.536~T_9^{6/3}+215.4007~T_9^{7/3}\Big] \nonumber\\
    +\frac{4.66187\times10^6}{T_9^{1/2}}\exp\Big(\frac{-10.92388}{T_9^{1/2}}\Big) \nonumber\\
  +\frac{8.5529\times10^7}{T_9}\exp\Big(\frac{-15.50687}{T_9^{1/3}}\Big)  \nonumber\\
  +\frac{16674.76}{T_9^{1/3}}\exp\Big(\frac{-7.86955}{T_9^{1/2}}\Big) \nonumber\\
     -\frac{77.74082}{T_9^2}\exp\Big(\frac{-1.38331}{T_9^2}\Big),  
\label{seqn6}
\end{eqnarray}
\vspace{0.0cm}

\noindent
\vspace{0.25cm} 
$^7$Li(n,$\gamma$)$^8$Li \cite{Du21} used for temperature range $\le 10T_9$
\begin{eqnarray}
N_A<\sigma v> =0.7306979\times10^4+ \nonumber\\
  +\frac{0.1116921\times10^7}{T_9^{5/2}}\exp\Big(\frac{-6.682375}{T_9^{1/2}}\Big) \nonumber\\
  -0.9260274~T_9^{2.871425},  
\label{seqn7}
\end{eqnarray}
\vspace{0.0cm}

\noindent
\vspace{0.25cm} 
$^{11}$B(n,$\gamma$)$^{12}$B \cite{Du20a} used for temperature range $\le 10T_9$
\begin{eqnarray}
N_A<\sigma v> =\frac{33.91692}{T_9^{2/3}}\exp\Big(\frac{-0.38836}{T_9^{1/3}}\Big)\times\Big[1.0+ \nonumber\\
+55.98557~T_9^{1/3}-211.765~T_9^{2/3}+611.5288~T_9 \nonumber\\
-717.9438~T_9^{4/3}+327.591~T_9^{5/3}\Big]
\label{seqn8}
\end{eqnarray}
\noindent
where the reaction rate $N_A<\sigma v>$ is expressed in units of cm$^3$mol$^{-1}$s$^{-1}$. Beyond the applicable temperature ranges mentioned above for each of the reaction rates, previously used \cite{Si19} reaction rates have been retained.     
       
\noindent 
\section{ Results and Discussion }
\label{section5}

    The impact of basic constants and nuclear reaction rates on primordial nucleosynthesis is thoroughly investigated. Using the most current experimental value for the neutron lifetime $877.75\pm0.28_{stat}+0.22/-0.16_{syst}$ s \cite{Go21} and the value of $\eta = \eta_{10} \times 10^{-10} = 6.0914 \pm 0.0438 \times 10^{-10}$ \cite{WMAP,WMAP1} for the baryon-to-photon ratio, all the computations previously described for the conventional BBN and with changed reaction rates for $^3$H(p,$\gamma$)$^4$He, $^6$Li(p,$\gamma$)$^7$Be, $^7$Be(p,$\gamma$)$^8$B, $^{13}$N(p,$\gamma$)$^{14}$O, $^7$Li(n,$\gamma$)$^8$Li and $^{11}$B(n,$\gamma$)$^{12}$B \cite{Du17a,Du22,Du19,Du20,Du21,Du20a} have been performed. The Table-I compares the findings of the current calculations with those of our earlier ones \cite{Si19,Mi12} and those of other recent studies \cite{Coc14,Coc15,Cy16,Il20,Pi21}. The experimental uncertainties in the values of $\tau_n$ and $\eta_{10}$ are the source of the theoretical uncertainties listed in the Table-I. However, the reaction rates are the second major source of uncertainty, and they would undoubtedly exacerbate the theoretical uncertainties mentioned in the present work. We discover that the most recent values of the basic constants and the new reaction rates result in a small decrease in the mass fraction of helium, relative abundances of deuteron, ${^3}$He and ${^7}$Li producing a minor improvement over the results previously obtained in conventional BBN calculations \cite{Si19}. 

    It was pointed out that the existence of metastable, lifetime $> 10^{3}$ s, negatively charged electroweak-scale particles might alter the predictions for lithium and other primordial elemental abundances for A $>4$ via the formation of bound states with nuclei \cite{Po07,Ka07} during BBN. It was also conjectured that hypothetical charged massive particles (CHAMPs), when present during the BBN era, may significantly alter the synthesis of light elements in comparison to a standard BBN scenario \cite{Je08,Ku08,Je08a}. The possible simultaneous solution of the $^{6,7}$Li problems for a relic decaying with lifetime of about $10^{3}$ s is only weakly dependent on the relic being neutral or charged, unless its hadronic branching ratio is small. It was shown that within CHAMP BBN, the existence of further parameter space for a simultaneous solution of the $^{6,7}$Li problems for long decay times $\geq 10^{6}$ s seems possible but fairly unlikely \cite{Je08a}. Concerning new physics beyond the standard model, the computer code AlterBBN \cite{Ar12,Ar20} delivers fast and reliable calculation of the Big-Bang nucleosynthesis constraints in alternative cosmologies. It does not rely on any closed external library or program for calculation of the abundances of the elements from BBN. The elemental abundances obtained using this code were ${^4}$He=0.2473$\pm$0.0003, D/H=(2.463$\pm$0.038)$\times 10^{-5}$, ${^3}$He/H=(1.034$\pm$0.016)$\times 10^{-5}$ and, most importantly, ${^7}$Li/H=(5.376$\pm$0.352)$\times 10^{-10}$. From Table-I, it may be ratiocinated that the results of the present calculations with standard BBN still provide slightly better estimates for the primordial elemental abundances.

    The effects of changes in neutron lifetime and nuclear reaction rates individually and together can be visualized through Table-II. If the value of neutron lifetime 880.3$\pm$1.1 s \cite{Ol14} used in our previous work \cite{Si19} is retained then there is no change in the mass fraction of helium, relative abundances of deuteron and ${^3}$He and only ${^7}$Li relative abundance becomes $4.427\times 10^{-10}$. Our calculations show that if the neutron lifetime is shortened by about 2.55 seconds (0.29$\%$) then there is decrease in the mass fraction of ${^4}$He (0.24$\%$) but for relative abundances of ${^7}$Li, it is even smaller (0.14$\%$) as evident from Table-II. The uncertainties of the calculated reaction rates arising due to errors associated with experimental data for the reactions $^3$H(p,$\gamma$)$^4$He, $^6$Li(p,$\gamma$)$^7$Be, $^7$Be(p,$\gamma$)$^8$B, $^{13}$N(p,$\gamma$)$^{14}$O, $^7$Li(n,$\gamma$)$^8$Li and $^{11}$B(n,$\gamma$)$^{12}$B are, respectively, 1$\%$, 5$\%$, 1$\%$, 5$\%$, 5$\%$ and 5$\%$ \cite{Du17a,Du22,Du19,Du20,Du21,Du20a}. The uncertainties in the theoretical calculations induced by the uncertainties of the calculated reaction rates have also been shown in the last column of Table-II. The theoretical estimates show that these alterations only slightly improve upon an earlier, significant fall in the abundance of ${^7}$Li  by $\sim 12\%$. Inferentially, the majority of these nuclear reactions have no impact on the primordial lithium abundance problem of BBN, even with large nuclear physics uncertainty.
    
\noindent
\section{ Summary and conclusion }
\label{section6}

    The primordial abundances depend on the astrophysical thermonuclear reaction rates and on three additional parameters, the number of light neutrino flavors, the neutron lifetime and the baryon-to photon ratio in the Universe. The theoretical predictions of the primordial abundances of elements in the Big-Bang Nucleosynthesis (BBN) are dominated by uncertainties in the Maxwellian-averaged thermonuclear nuclear reaction rates. The effect of modifying some of these reaction rates on elemental abundance in BBN has been investigated. We previously looked into the impact of changing thirty-five thermonuclear reaction rates on the yields of light element abundances in BBN \cite{Mi12}. Later, the neutron lifetime and the baryon-to-photon ratio were also replaced by the latest values available and the reaction rates for d($^4$He,$\gamma$)$^6$Li, t($^4$He,$\gamma$)$^7$Li and ${^3}$He(${^4}$He,$\gamma$)${^7}$Be (latter of which is used directly to estimate the formation of ${^7}$Li as a result of decay by electron capture) \cite{Du17} were modified as well. The primordial light element abundance yields as functions of temperature or evolution time were studied \cite{Si19}. In the present work, calculations are performed using new reaction rates for $^3$H(p,$\gamma$)$^4$He, $^6$Li(p,$\gamma$)$^7$Be, $^7$Be(p,$\gamma$)$^8$B, $^{13}$N(p,$\gamma$)$^{14}$O, $^7$Li(n,$\gamma$)$^8$Li and $^{11}$B(n,$\gamma$)$^{12}$B along with the most recent experimentally measured neutron lifetime value of $877.75\pm0.28_{stat}+0.22/-0.16_{syst}$ s \cite{Go21}. We observe from theoretical calculations that these changes result in marginal improvement over a sizable twelve percent reduction in the abundance of $^7$Li achieved earlier \cite{Si19}. Adding new reaction rates to the BBN code and expanding the reaction network were found to have almost little impact on the BBN abundances in a few other recent investigations \cite{Coc14,Coc15,Cy16,Bo10}. It is important to note that a later study \cite{Sb10} revised the previously measured \cite{Ho09} relative abundance of $^7$Li ($1.1\pm0.1 \times 10^{-10}$) upward by roughly 44 percent. Although the theoretical and the observed values appear to be converging, the lowest limit of current theoretical estimate is still larger by a factor of 2.25 compared to the highest limit of the observed value of $^7$Li relative abundance. It is important to mention that there may be further scope of improvement and that the present observational capabilities do not yet allow the detection of primeval Lithium in very metal-poor stars of the galactic halo.

\begin{acknowledgments}

    One of the authors (DNB) acknowledges support from Science and Engineering Research Board, Department of Science and Technology, Government of India, through Grant No. CRG/2021/007333.

\end{acknowledgments}


\vspace {0.0cm}
\noindent
\begin{thebibliography}{99}

\bibitem{Ho64} F. Hoyle and R. J. Tayler, Nature (London) {\bf 203}, 1108 (1964).

\bibitem{Planck} P. A. R. Ade et al. (Planck Collaboration XVI), Astron. Astrophys. {\bf 571}, A16 (2014).

\bibitem{Planck1} P. A. R. Ade et al. (Planck Collaboration XIII), Astron. Astrophys. {\bf 594}, A13 (2016).

\bibitem{WMAP} E. Komatsu et al., Astrophys. J. Suppl. {\bf 192}, 18 (2011). 

\bibitem{WMAP1} G. Hinshaw et al., Astrophys. J. Suppl. {\bf 208}, 19 (2013).

\bibitem{Di82} D. Dicus, E. Kolb, A. Gleeson, E. Sudarshan, V. Teplitz and M. Turner, Phys. Rev.
D 26 (1982) 2694.

\bibitem{Ol14} K. A. Olive et al. (Particle Data Group), Chin. Phys. {\bf C 38}, 090001 (2014)  URL: http://pdg.lbl.gov.

\bibitem{Si19} V. Singh, J. Lahiri, D. Bhowmick and D. N. Basu, Journal of Experimental and Theoretical Physics {\bf 128}, 707 (2019).  
  
\bibitem{Wi11} F. Wietfeldt and G. Greene, Rev. Mod. Phys. {\bf 83}, 1173 (2011).

\bibitem{Yo14} A. R. Young et al., J. Phys. {\bf G 41}, 114007 (2014).

\bibitem{Go21} F. M. Gonzalez et al., Phys. Rev. Lett. {\bf 127}, 162501 (2021).

\bibitem{Ma05} G. J. Mathews, T. Kajino and T. Shima Phys. Rev. {\bf D 71}, 021302(R) (2005).

\bibitem{No00} K. M. Nollett and S. Burles, Phys. Rev. {\bf D 61}, 123505 (2000).

\bibitem{Cy02} R. H. Cyburt, B. D. Fields, and K. A. Olive, Astropart. Phys. {\bf 17}, 87 (2002).

\bibitem{Cy04} R. H. Cyburt, Phys. Rev. {\bf D 70}, 023505 (2004).
 
\bibitem{Cy08} R. H. Cyburt, B. D. Fields and K. Olive, J. Cosm. Astropart. Phys. {\bf 11}, 12 (2008).

\bibitem{Se04} P. D. Serpico, S. Esposito, F. Iocco, G. Mangano, G. Miele, and O. Pisanti, J. Cosmol. Astropart. Phys. {\bf 12}, 010 (2004).

\bibitem{Io09} F. Iocco, G. Mangano, G. Miele, O. Pisanti, and P. D. Serpico, Phys. Rep. {\bf 472}, 1 (2009).

\bibitem{Fu10} G. M. Fuller and C. J. Smith, Phys. Rev. {\bf D 82}, 125017 (2010).

\bibitem{Fo88} G. R. Caughlan, and W. A. Fowler, Atom. Data Nucl. Data Tables {\bf 40}, 283 (1988).

\bibitem{An99} C. Angulo et al., Nucl. Phys. {\bf A 656}, 3 (1999).

\bibitem{Mi12} Abhishek Mishra and D. N. Basu, Rom. J. Phys. {\bf 57}, 1317 (2012).

\bibitem{Du17} S. B. Dubovichenko, A. V. Dzhazairov-Kakhramanov and N. A. Burkova, arXiv:1706.05245;\\ S. B. Dubovichenko, Russ. Phys. J. {\bf 60}, 1143 (2017).

\bibitem{Bo08} R. N. Boyd, {\it An Introduction to Nuclear Astrophysics} (University of Chicago, Chicago, 2008), 1st ed.

\bibitem{Bl55} J. M. Blatt and V. F. Weisskopf, {\it Theoretical Nuclear Physics} (John Wiley $\&$ Sons, New York; Chapman $\&$ Hall Limited, London.) 

\bibitem{Mu10} Tapan Mukhopadhyay, Joydev Lahiri and D. N. Basu, Phys. Rev. {\bf C 82}, 044613 (2010); {\it ibid} Phys. Rev. {\bf C 83}, 039902(E) (2011).

\bibitem{Aver15} E. Aver, K. A. Olive and E. D. Skillman, J. Cosmology Astropart. Phys. {\bf 07}, 011 (2015); E. Aver, D. A. Berg, K. A. Olive, R. W. Pogge, J. J. Salzer and E. D. Skillman, 2020, arXiv:2010.04180.

\bibitem{Cooke14} R. Cooke, M. Pettini, R. A. Jorgenson, M. T. Murphy and C. C. Steidel, Astrophys. J. {\bf 781}, 31 (2014); R. J. Cooke, M. Pettini and C. C. Steidel, Astrophys. J. {\bf 855}, 102 (2018).

\bibitem{Ba02} T. Bania, R. Rood and D. Balser, Nature {\bf 415}, 54 (2002).

\bibitem{Sp82} F. Spite and M. Spite, Astron. Astrophys. {\bf 115}, 357 (1982).

\bibitem{Sb10} L. Sbordone, P. Bonifacio, E. Caffau et al., Astron. Astrophys. {\bf 522}, A26 (2010).

\bibitem{Sm93} M.S. Smith, L.H. Kawano and R.A. Malaney, Astrophys. J. Suppl. {\bf 85}, 219 (1993).

\bibitem{Wa67} R. Wagoner, W. A. Fowler, and F. Hoyle, Astrophys. J. {\bf 148}, 3 (1967).

\bibitem{Wa69} R. Wagoner, Astrophys. J. Supp. {\bf 18}, 247 (1969).

\bibitem{Ka92} L. Kawano, FERMILAB Report No. PUB-92/04-A, January 1992 (unpublished).

\bibitem{An04} P. Descouvemont, A. Adahchour, C. Angulo, A. Coc, E. Vangioni-Flam, Atom. Data Nucl. Data Tables {\bf 88}, 203 (2004).

\bibitem{Coc14} A. Coc, J.-P. Uzan and E. Vangioni, J. Cosmology Astropart. Phys. {\bf 10}, 050 (2014).

\bibitem{Coc15} A. Coc, P. Petitjean, J.-P. Uzan, E. Vangioni, P. Descouvemont, C. Iliadis and R. Longland, Phys. Rev. {\bf D 92}, 123526 (2015).

\bibitem{Cy16} R. H. Cyburt, B. D. Fields, K. A. Olive and T.-H. Yeh, Rev. Mod. Phys. {\bf 88}, 015004 (2016).

\bibitem{Il20} Christian Iliadis and Alain Coc, Astrophys. J. {\bf 901}, 127 (2020).

\bibitem{Pi21} Cyril Pitrou, Alain Coc, Jean-Philippe Uzan and Elisabeth Vangioni, Mon. Not. R. Astron. Soc. {\bf 502}, 2474 (2021). 

\bibitem{Du17a} S. B. Dubovichenko, A. V. Dzhazairov-Kakhramanova and N. V. Afanasyeva, Nucl. Phys. {\bf A 963}, 52 (2017).

\bibitem{Du22} S. B. Dubovichenko, A. S. Tkachenko, R. Ya. Kezerashvili, N. A. Burkova and A. V. Dzhazairov-Kakhramanov, Phys. Rev. {\bf C 105}, 065806 (2022).

\bibitem{Du19} S. B. Dubovichenko, N. A. Burkova, A. V. Dzhazairov-Kakhramanov and A. S. Tkachenko, Nucl. Phys. {\bf A 983}, 175 (2019).

\bibitem{Du20} S. B. Dubovichenko, R. Ya. Kezerashvili, N. A. Burkova, A. V. Dzhazairov-Kakhramanov and B. Beisenov, Phys. Rev. {\bf C 102}, 045805 (2020).

\bibitem{Du21} N. A. Burkova, S. B. Dubovichenko, A. V. Dzhazairov-Kakhramanov and S. Zh. Nurakhmetova, J. Phys. {\bf G 48}, 045201 (2021).

\bibitem{Du20a} S. B. Dubovichenko, N. A. Burkova, A. V. Dzhazairov-Kakhramanov and A. S. Tkachenko, Astropart. Phys. {\bf 123}, 102481 (2020).

\bibitem{Po07} M. Pospelov, Phys. Rev. Lett. {\bf 98}, 231301 (2007).

\bibitem{Ka07} M. Kawasaki, K. Kohri and T. Moroi, Phys. Lett. {\bf B 649}, 436 (2007).

\bibitem{Je08} K. Jedamzik, JCAP {\bf 0803}, 008 (2008).

\bibitem{Ku08} M. Kusakabe, T. Kajino, R. N. Boyd and T. Yoshida, Astrophys. J. {\bf 680}, 846 (2008).
 
\bibitem{Je08a} K. Jedamzik, Phys. Rev. {\bf D 77}, 063524 (2008).

\bibitem{Ar12} A. Arbey, Comput. Phys. Commun. {\bf 183}, 1822 (2012). 

\bibitem{Ar20} A. Arbey, J. Auffinger, K. Hickerson and E. Jenssen, Comput. Phys. Commun. {\bf 248}, 106982 (2020).

\bibitem{Bo10} R. N. Boyd, C. R. Brune, G. M. Fuller and C. J. Smith, Phys. Rev. {\bf D 82}, 105005 (2010).

\bibitem{Ho09} A. Hosford, S. G. Ryan, A. E. Garcia-Perez, J. E. Norris and K. A. Olive, Astron. Astrophys. {\bf 493}, 601 (2009).

\end{thebibliography}
\end{document}